\documentstyle{elsart}

\begin{document}
\begin{frontmatter}

\address{~~~~~~~~~~~~~~~~~~~~~~~~~~~~~~~~~~~~~~~~~~~~~~~~~~~~~
~~~~~~~~~~~~~~SISSA 119/98/EP} 
\address{~~~~~~~~~~~~~~~~~~~~~~~~~~~~~~~~~~~~~~~~~~~~~~~~~~~~~
~~~to be published in Phys. Lett. B}

\title{Double D-term inflation}
\author{Julien Lesgourgues$^{1}$}
\address{\it SISSA/ISAS, Via Beirut 4, 34014 Trieste,Italy}

\date{1 March 1999}
\maketitle

\begin{abstract}
Comparisons of cosmological models to current data show that the presence
of a non-trivial feature in the primordial power spectrum of fluctuations,
around the scale $k \sim 0.05 h$Mpc$^{-1}$, is an open
and exciting possibility, testable in a near future.
This could set new constraints on inflationary models.
In particular, current data favour a $\Lambda$CDM model with a steplike 
spectrum, and more power on small scales.
So far, this possibility has been
implemented only in toy models of inflation. 
In this work, we propose a supersymmetric model 
with two $U(1)$ gauge symmetries, associated with two Fayet-Iliopoulos terms. 
Partial cancellation of the Fayet-Iliopoulos by one of the 
charged fields generates a step in the primordial power spectrum of 
adiabatic perturbations. We show that when this field if charged under both
symmetries, the spectrum may have more power on small scales.
\end{abstract}


\end{frontmatter}

\section{Introduction}
A most exciting aspect of inflationary models is the possibility of 
constraining them through observations of cosmological perturbations. 
Generally, the power spectrum of primordial fluctuations can be easily related
to the inflation potential \cite{LL}. In more complicated models, 
like multiple-stage inflation \cite{KLS,RS,ST}, 
slow-roll conditions may be violated; then, the primordial
spectrum has to be derived from first principles, and exhibits 
a non-trivial feature at some scale. 
These scenarios shouldn't be regarded as unrealistic: 
inflation has to take place within the framework 
of a high-energy theory, with many scalar fields. 
So, the real question is not to know wether multiple-stage inflation
can occur, but wether it is predictive, or
just speculative. Multiple-stage inflation has already
been invoked for many purposes :
generation of features in the power spectrum
at observable scales, setting of initial conditions,
description of the end of inflation. Here we are interested in the
first possibility. The range of cosmological perturbations observable today
in the linear or quasi-linear regime has been generated during approximately
8 {\it e}-folds, 50 {\it e}-folds before the end of inflation.
So, a feature in the primordial power spectrum could very well be observable, 
without any fine-tuning;
only observations can rule out this possibility.

In the radiation and matter dominated Universe, primordial fluctuations of
the metric perturbations couple with all components, leading to the formation
of Cosmic Microwave Background (CMB) anisotropies and Large Scale Structure
(LSS). It is well established that a pure Cold Dark Matter (CDM) scenario,
in a flat universe, and with a scale-invariant or even a tilted primordial
spectrum, is in conflict with LSS data. So, many variants of this scenario
have been considered, including a Broken Scale Invariant (BSI) power spectrum.
Indeed, standard CDM predictions are improved when {\it less} 
power is predicted on small scales. 
Specific cases have been compared accurately with both
LSS data and recent CMB experiments,
including the double-inflationary spectrum of \cite{P94}, and 
the steplike spectrum of \cite{S92}. 
However, even with such spectra, it has been shown \cite{LP,LPS1,LPS2} 
that standard CDM cannot be made compatible with all observations.
Independently, recent constraints from Ia supernovae \cite{PGRW} 
strongly favour the $\Lambda$CDM scenario, with a cosmological constant 
$\Lambda \simeq 0.6-0.7$ and a Hubble parameter $h=0.6-0.7$. 
In this framework, a reasonable fit to all CMB and LSS data can be obtained 
with a flat or a slighly tilted spectrum. So, $\Lambda$CDM is very promising
and should be probably considered as a current standard picture, 
from which possible
deviations can be tested. In this respect, in spite of large error bars, 
some data indicate a possible sharp feature in the primordial power 
spectrum around the  scale $k\simeq 0.05 h$Mpc$^{-1}$. 
First, APM redshift survey data seem to point out
the existence of a step in the present matter power spectrum \cite{GB,RS}. 
Second, Abell galaxy cluster data exhibit a similar feature 
\cite{E} (at first sigh, it looks
more like a spike than like a step, but in fact a steplike primordial spectrum
multiplied with the $\Lambda$CDM transfer function reproduces this shape 
\cite{LPS1}). 
Third, this scale corresponds to $l \sim 300$, {\it i.e.} to the
first accoustic peak in the CMB anisotropies, and increasing power
at this scale, through a bump or a step, would lead to a better agreement
with Saskatoon. In the next years, future observations will either
rule out this possibility 
(which will then be attributed to underestimated errorbars),
or confirm the existence of a feature, and
precisely constraint its shape. 
In \cite{LPS1,LPS2}, we compared a specific BSI $\Lambda$CDM model 
with CMB and LSS data.
The primordial spectrum was taken from a toy model proposed by Starobinsky 
\cite{S92}, in which
the derivative of the inflaton potential changes very rapidly in a narrow 
region. In this case, the primordial spectrum 
consists in two approximately flat plateaus, separated by a step. This
model improves the fits to CMB and LSS data (even when the cluster data
are not taken into account). It also reproduces fairly well the feature of
Einasto et al. \cite{E} when the step is inverted with respect to the
usual picture, {\it i.e.}, with {\it more} power on small scales. 
Independently, in a preliminary work \cite{GAS} (motivated
by the inflationary model of \cite{BAFO}),
a primordial spectrum with a bump centered
at $k\simeq 0.06 h$Mpc$^{-1}$ was compared with CMB
and LSS data, including more redshift surveys than in \cite{LPS1}, 
and not taking Einasto et al. \cite{E} data into account. 
It is remarkable that among many cosmological scenarios,
this BSI spectrum combined with $\Lambda$CDM yields one of the best fits.

So, we have good motivations for searching inflationary models based on 
realistic high-energy physics, that predict a bump or an inverted
step at some scale, and approximately scale-invariant plateaus far from it.
Successfull comparison with the data requires the deviation
from scale invariance to be concentrated in a narrow region 
$k_1 \leq k \leq k_2$; roughly, $k_2/k_1$ should be in the range 2-10. 
This is quite challenging, because
double inflationary models studied so far predict systematically
{\it less} power on small scale, with  
a logarithmic decrease on large scale rather than a plateau. 
However, these models were based on the general framework of chaotic
inflation. Today, the best theoretically motivated framework
is hybrid inflation \cite{LLL}. Indeed, hybrid inflation has many attractive 
properties, and appears naturally in supersymmetric models: 
the inflaton field(s) follow(s) one of the flat directions of the potential,
and the approximately constant potential energy density is provided by the
susy-breaking F or D-term. When the F-term does not vanish, 
conditions for inflation are generally spoiled by supergravity
corrections, unless interesting but rather complicated variants are considered
(for a very good review, see \cite{LR}). On the contrary, the simple D-term
inflation mechanism proposed by Bin\'etruy \& Dvali \cite{BD} and Haylo
\cite{HA} is naturally
protected against supergravity corrections, and can be easily implemented
in realistic particle physics models, like the GUT model of reference
\cite{DR}, without
invoking {\it ad hoc} additional sectors.   
If supergravity is considered as an effective theory from superstrings, 
D-term inflation is hardly compatible with heterotic
strings theory \cite{LR,ERR}, but consistent with type I string theory
\cite{HA2}.

Our goal is to show that in this framework, a simple mechanism
can generate a steplike power spectrum with {\it more} power on small scales.
This mechanism is based on a variant of D-term inflation, with two 
Fayet-Iliopoulos terms. However, the fact that it is D-term inflation 
is not crucial for our purpose: 
a similar lagrangian could be obtained with two non-vanishing F-terms,
or one D-term plus one F-term, like in \cite{ST}. We will not consider the
link between the model of this paper and string theory, putting by hand 
the Fayet-Iliopoulos terms from the beginning.
In a very interesting paper, Tetradis \& Sakellariadou \cite{ST}
studied a supersymmetric double inflationary model with a quite similar 
lagrangian. However, the motivation of these authors is
to save standard CDM. So, they are pushed to 
regions in parameter space quite different from us, and do not
consider a steplike spectrum with flat plateaus, but rather
a power-law spectrum with $n=1$ on large scales and $n<1$ on small scales. 

\section {The model}

We consider a supersymmetric model with two gauge symmetries 
$U(1)_A \times U(1)_B$, and two associated
Fayet-Iliopoulos positive terms $\xi_A$ and $\xi_B$
(there is no motivation from string theory to do so, at least at the moment,
but SUSY and SUGRA allow an arbitrary number of Fayet-Iliopoulos
terms to be put by hand in the lagrangian).
In this framework, the most simple workable model involves six
scalar fields: two singlets $A$ and $B$, and four charged fields
$A_{\pm}$, $B_{\pm}$, with charges $(\pm 1,0)$, $(\pm 1,\pm 1)$.  
Let us comment this particular choice. First,
the presence of two singlets is crucial. With only one singlet
coupling to both $A_\pm$ and $B_\pm$, we would still have double-inflation, 
but the second stage would be driven by both F and D-terms, 
and no sharp feature would be predicted in the primordial spectrum.
Second, each charged field could be charged under one symmetry only;
then, a steplike spectrum would be generated, but with necessarily 
{\it less} power
on small scales. Here, the fact that $B_-$ has a charge $-1$
under both symmetries is directly responsible for the inverted step,
as will become clear later. Finally,
both global susy and supergravity versions of this model can be studied:
supergravity corrections would change the details
of the scenario described thereafter, but not its main features.
We consider the superpotential:
$
W=\alpha A A_+ A_- + \beta B B_+ B_-
$
(with a redefinition of $A$ and $B$, we have suppressed 
terms in $B A_+ A_-$, $A B_+ B_-$, and made ($\alpha$, $\beta$) real 
and positive).
In global susy, the corresponding tree-level scalar potential is:
\begin{eqnarray} 
&V&= \alpha^2 |A|^2 (|A_+|^2 \!\!\! + \! |A_-|^2) + \alpha^2 |A_+ A_-|^2
+ \frac{g_A^2}{2} (|A_+|^2 \!\!\! - \! |A_-|^2 \!\!\! + \! |B_+|^2 \!\!\! 
- \! |B_-|^2 \!\!\! + \xi_A)^2
\nonumber \\
&+& \beta^2 |B|^2 (|B_+|^2 \!\!\! + \! |B_-|^2) + \beta^2 |B_+ B_-|^2
+ \frac{g_B^2}{2} (|B_+|^2 \!\!\! - \! |B_-|^2 \!\!\! + \xi_B)^2, \label{vtree}
\end{eqnarray}
with a global supersymmetric vacuum in which all fields are at the origin,
excepted $|B_-|=\sqrt{\xi_B}$, and, depending on the sign of $(\xi_A-\xi_B)$,
$|A_-|$ or $|A_+|=\sqrt{|\xi_A-\xi_B|}$.

\section{The two slow-roll inflationary stages}

\subsection{Inflationary effective potential}

There will be generically two stages of inflation, 
provided that initial conditions for $(A,B)$ obey to :
\begin{equation} \label{condition}
|A|^2 \geq \frac{g_A^2 \xi_A}{\alpha^2}, \qquad 
|B|^2 \geq \frac{g_A^2 \xi_A + g_B^2 \xi_B}{\beta^2}.
\end{equation}
Then, charged fields have positive squared masses
and stand in their (local) minimum
$A_\pm=B_\pm=0$ (for a discussion of the charged fields initial conditions, 
see for instance \cite{LT,BCT}). 
$A$ and $B$ have a constant phase, while their moduli 
$\tilde{A} \equiv |A|/\sqrt{2}$ and $\tilde{B} \equiv |B|/\sqrt{2}$ behave
like two real inflaton fields and roll to the origin,
until one inequality in (\ref{condition})
breaks down. We assume that the condition on $B$ breaks first.

During this first stage, the field evolution is driven by the effective
potential:
$V_1= (g_A^2 \xi_A^2 + g_B^2 \xi_B^2)/2 + \Delta V_1$. 
The one-loop correction $\Delta V_1$ is small 
($\Delta V_1 \ll V_1$) \cite{HA2}, 
but crucial for the field evolution. It consists in two terms with a 
logarithmic dependence on $A$ and $B$. The former takes a simple form
following from \cite{DSS}, 
because we can assume $g_A^2 \xi_A \ll \alpha^2 |A|^2 $.
The latter is more complicated because the dimensionless quantity 
$b \equiv (g_A^2 \xi_A + g_B^2 \xi_B)/(\beta^2 |B|^2)$ 
goes to one when $B$ reaches its critical value. 
The exact expressions are:
\begin{eqnarray}
&\Delta V_1& = 
\frac{g_A^4 \xi_A^2}{32 \pi^2} \left( \ln \frac{\alpha^2 |A|^2}{\Lambda^2}
+ \frac{3}{2} \right) \\
&+& \frac{(g_A^2 \xi_A + g_B^2 \xi_B)^2}{32 \pi^2} \left( 
\ln \frac{\beta^2 |B|^2}{\Lambda^2} + \frac{(1+b)^2 \ln (1+b)
+ (1-b)^2 \ln (1-b)}{2b^2} \right) \nonumber
\end{eqnarray}
($\Lambda$ is the renormalization energy scale at which $g_A$ and $g_B$ must
be evaluated).
When $b \ll 1$, the term involving $b$ tends to $\frac{3}{2}$. 
Even at $b=1$, this term only
increases the derivative $(\partial V_1 / \partial \tilde{B})$ 
by a factor $2 \ln 2$ : so, it can be 
neglected in qualitative studies. In this approximation, it is easy to
calculate the trajectories of $A$ and $B$, and to note that $B$ reaches its
critical value before $A$ only if the initial field values
obey to:
\begin{equation} \label{ini} 
\frac{|B|_0}{|A|_0} < 1+\frac{g_B^2 \xi_B}{g_A^2 \xi_A}.
\end{equation}
This condition is natural, in the sense that it allows
$|A|_0$ and $|B|_0$ to be of the same order of magnitude, whatever the values
of $g_{A,B}$ and $\xi_{A,B}$.

At the end of the first stage, ($B$, $B_-$) evolve to another false vacuum:
$B=0$, $|B_-|^2=(g_A^2 \xi_A + g_B^2 \xi_B)/(g_A^2 + g_B^2)$. 
During this transition,
the charged fields $B_+$, $A_\pm$ remain automatically stable 
if we impose $\xi_B \leq 2 \xi_A$.
Afterwards, a second stage occurs: $A$ rolls to the origin, driven
by the potential: 
\begin{equation}
V_2 = \frac{g_A^2 g_B^2 (\xi_A-\xi_B)^2}{2(g_A^2 + g_B^2)} \left( 1 +
\frac{g_A^2 g_B^2}{16 \pi^2 (g_A^2 + g_B^2)} 
\left( \ln \frac{\alpha^2 |A|^2}{\Lambda^2} + \frac{3}{2} \right) \right),
\end{equation}
until $|A_+|$ or $|A_-|$ becomes unstable, and quickly drives the fields 
to the supersymmetric minimum.

\subsection{Second stage of single-field inflation}

Let us focus first on the second stage of inflation,
in order to find the small scale primordial power spectrum ({\it i.e.}, if 
the second stage starts at $t=t_2$, and $k_2 \equiv a(t_2) H(t_2)$,
the power spectrum at scales $k>k_2$). This stage should
last approximately $N \simeq 50$ {\it e}-folds, so that the transition 
takes place when scales observable today cross the Hubble radius.

A standard calculation shows that for
$\alpha$ of order one, the second slow-roll condition breaks
before $A$ reaches its critical value (which is given by
$\alpha^2 |A_{C}|^2  = \frac{g_A^2 g_B^2}{g_A^2+g_B^2} 
|\xi_A-\xi_B|$). Integrating back in time, 
we find that $N$ {\it e}-folds before the end of inflation, 
\begin{equation} \label{abreak}
|A|= \sqrt{\frac{N}{2 \pi^2}} \frac{g_A g_B}
{\sqrt{g_A^2 + g_B^2}} {\mathrm M}_P
\end{equation}
(we are using the reduced Planck mass 
${\mathrm M}_P \equiv (8 \pi G)^{-1/2} = 2.4 \times 10^{18} {\mathrm GeV}$).
Then, the primordial spectrum can be easily calculated, using the single-field
slow-roll expression \cite{LL}:
\begin{equation} \label{secspec}
\delta_H^2 =  \frac{1}{75 \pi^2 {\mathrm M}_P^6} 
\frac{V_2^3}{(d V_2 / d \tilde{A})^2} = 
\frac{16 \pi^2}{75 {\mathrm M}_P^6}
\frac{g_A^2 + g_B^2}{g_A^2 g_B^2}
(\xi_A - \xi_B)^2 |A|^2.
\end{equation}
To normalize precisely this spectrum (\ref{secspec}) 
to COBE, it would be necessary to
calculate the contribution of cosmic strings generated by symmetry breaking
\cite{J}, to make assumptions about the geometry and matter content of the
universe, and to fix the amplitude of the step in the primordial
power spectrum (between COBE scale and $k_2$). However, if
perturbations generated by cosmic strings are not dominant, and if
the primordial spectrum is approximately scale-invariant
as required by observation, we can estimate the order of magnitude of the
primordial power spectrum at all observable scales: $\delta_H^2 
\sim 5 \times 10^{-10}$. So, at $k=k_2$, inserting
(\ref{abreak}) into (\ref{secspec}), we find the constraint:
\begin{equation} \label{norm}
\sqrt{|\xi_A-\xi_B|} \sim 3 \times 10^{-3} {\mathrm M}_P
\sim 10^{15-16} {\mathrm GeV}.
\end{equation}
At first sight, this constraint
could be satisfied when ($\sqrt{\xi_A}$, $\sqrt{\xi_B}$) are both much 
greater than $10^{-3} {\mathrm M}_P$, and very close to each other.
Then, however, the amplitude of the large-scale plateau would violate the
COBE normalization, as can be seen from the following. 
Also, there is no reason to believe that one term is much smaller than the 
other: this would raise a fine-tuning problem. So, we will go on assuming
that both Fayet-Iliopoulos terms have an order of magnitude 
$\sqrt{\xi_A}\sim \sqrt{\xi_B} \sim 10^{-3} {\mathrm M}_P$, just as
in single D-term inflation.

The spectrum tilt $n_S$ at $k=k_2$ can be deduced from the  
slow-roll parameters ($\epsilon$, $\eta$) \cite{LL}. 
Like in any model of single-field
D-term inflation, $\epsilon$ can be neglected, and
$n_S(k_2)=1+2\eta(k_2)=1-1/N \simeq 0.98$. So, the spectrum on small scales 
is approximately scale-invariant.

\subsection{First stage of two-field inflation}

During the first inflation,
the primordial spectrum calculation must be done carefully. 
If slow-roll conditions were to hold during
the transition between both inflationary stages, the evolution of metric 
perturbations (for modes outside the Hubble radius)
would be described at first order by the well-known slow-roll solution
(see for instance \cite{PS}):
\begin{equation} \label{phisr}
\Phi = - C_1 \frac{\dot{H}}{H^2}-
H \frac{d}{dt} \left( \frac{d_A V_A + d_B V_B}{V_A + V_B} \right),
\end{equation}
where $\Phi$ is the gravitational potential in the longitudinal gauge.
Here, $C_1$ is the time-independent coefficient of the growing
adiabatic mode, while $d_A$ and $d_B$ are coefficients related to 
the non-decaying
isocurvature mode (in fact, only $d_A-d_B$ is meaningful \cite{PS}).
The expression of $V_A$ and $V_B$ at a given time can be calculated
only if the whole field evolution is known, between
the first Hubble crossing and the end of inflation.
Formally, in the general case of multiple fields $\phi_i$, the $V_i$'s can be
found by integrating $d V_i = (\partial V / \partial \phi_i) d \phi_i$
back in time, starting from the end of inflation. This just means here
that during the second slow-roll, $V_A=V_2$, $V_B=0$, and during the first
slow-roll:
\begin{eqnarray}
V_A &=& \frac{g_A^2 g_B^2 (\xi_A - \xi_B)^2}{2 (g_A^2+g_B^2)}
+\frac{g_A^4 \xi_A^2}{32 \pi^2} \left( \ln \frac{\alpha^2 |A|^2}{\Lambda^2}
+ \frac{3}{2} \right) , \nonumber \\
V_B &=& \frac{( g_A^2 \xi_A + g_B^2 \xi_B )^2}{2 (g_A^2 + g_B^2)} 
 +\frac{(g_A^2 \xi_A + g_B^2 \xi_B)^2}{32 \pi^2} \left(
\ln \frac{\beta^2 |B|^2}{\Lambda^2} + \frac{3}{2} \right). 
\end{eqnarray}
We see immediately from (\ref{phisr}) with $V_B=0$ that the
isocurvature mode is suppressed during the second inflationary stage. 
On the other hand, it must be taken into account
during the first stage. This leads to the well-known expression
\cite{PS}:
\begin{equation} \label{lsp}
\delta_H^2 (k) = \frac{V}{75 \pi^2 {\mathrm M}_P^6} 
\left( \frac{V_{A}^2}{(d V_A / d \tilde{A})^2} + 
       \frac{V_{B}^2}{(d V_B / d \tilde{B})^2} \right)_k,
\end{equation} 
where the subscript $k$ indicates that quantities are evaluated at the
time of Hubble radius crossing during the first stage.

If slow-roll is briefly disrupted during the transition (this is the
interesting case if we want to generate a narrow step or bump),
the solution (\ref{phisr}) doesn't hold at any time, but we still 
have a more general exact solution, describing the adiabatic mode
(in the long-wavelenght regime):
\begin{equation} \label{phinsr}
\Phi = C_1 \left( 1-\frac{H}{a} \int_0^t a dt \right) + 
({\rm other~modes}).
\end{equation}
If, during the transition, the other modes are not dominant (as usually
expected), the matching of the three solutions : (\ref{phisr}) before the
transition, (\ref{phinsr}) during the transtion, and (\ref{phisr}) afterwards,
shows that $C_1$ is really the same number during all stages: 
the slow-roll disruption doesn't leave a signature on 
the large-scale power spectrum (\ref{lsp}).
On the other hand, if a specific phenomenon amplifies significantly
isocurvature modes during the transition, the same matching
shows that during the second stage, there will be an additional term
contributing to the adiabatic mode. This role may be played by
parametric amplification of metric perturbations, caused by 
oscillations of $B_-$. Generally speaking, the possibility
that parametric resonance could affect modes well outside the Hubble
radius is still unclear, and might be important in multi-field
inflation \cite{PARA}. 
In our case, this problem would require a carefull numerical 
integration, and would crucially depend on the details of the
one-loop effective potential during the transition. So, we leave this issue
for another publication, and go on under the standard assumption that 
expression (\ref{lsp}) can be applied to
any mode that is well outside the Hubble radius during the transition.

Let us apply this assumption to our model, and find
the large scale primordial power spectrum ({\it i.e.}, if 
the first stage ends at $t=t_1$, and $k_1 \equiv a(t_1) H(t_1)$,
the power spectrum at scales $k<k_1$). The contribution to $\delta_H$ arising
from perturbations in $A$ reads:
\begin{equation} \label{dha}
\delta_H^2 |_A \equiv \frac{V}{75 \pi^2 {\mathrm M}_P^6} 
\frac{V_A^2}{(d V_A / d \tilde{A})^2} 
= \frac{16 \pi^2}{75 {\mathrm M}_P^6} 
\frac{g_B^4 (g_A^2 \xi_A^2 + g_B^2 \xi_B^2) (\xi_A - \xi_B)^4}
{g_A^4 (g_A^2 + g_B^2)^2 \xi_A^4} |A|^2.
\end{equation}
The transition lasts approximately 1 {\it e}-fold (indeed, during this
stage, the evolution
is governed by second-order differential equations with damping terms
$+3H \dot{\phi_i}$, and $(B, B_-)$ stabilize within a time-scale
$\Delta t \sim H^{-1}$). During that time, $A$ is still in slow-roll
and remains approximately constant. So, using eqs. (\ref{secspec}) and 
(\ref{dha}),
it is straightforward to estimate the amplitude of the step in the 
primordial spectrum, under the assumption that $\delta_H (k_2) |_B$ is
negligible:
\begin{equation} \label{p}
p^2 \equiv \frac{\delta_H^2 (k_1) |_A}{\delta_H^2 (k_2)}
= \frac{(1-\xi_B/\xi_A)^2 (1+g_B^2 \xi_B^2/g_A^2 \xi_A^2)}{(1+g_A^2/g_B^2)^3}.
\end{equation}
Since we already imposed $\xi_B \leq 2\xi_A$, $p$ can easily be smaller
than one, so that we obtain, as desired, more power on small scales.
The simple explanation is that with $B_-$ charged under both symmetries, the
transition affects not only the dynamics of $(B, B_-)$, but also the
one-loop correction proportional to $(\ln |A|^2)$, in such way that the
slope $(\partial V / \partial \tilde{A})$ can decrease by the above factor $p$.

However, perturbations in $B$ must also be taken into account.
Their contribution to the large scale primordial spectrum reads:
\begin{equation}
\delta_H^2|_B \equiv \frac{V}{75 \pi^2 {\mathrm M}_P^4}
\frac{V_B^2}{(d V_B/ d \tilde{B})^2} = \frac{16 \pi^2}{75 {\mathrm M}_P^6} 
\frac{(g_A^2 \xi_A^2 + g_B^2 \xi_B^2)}
{(g_A^2+g_B^2)^2} |B|^2.
\end{equation}
At the end of the first stage, the value of $|B|$ is roughly given by
${\mathrm M}_P^2 (\partial^2 V_1 / \partial \tilde{B}^2) = V_1$, 
since after the breaking
of this slow-roll condition, eq. (\ref{lsp}) is not valid anymore, 
and $B$ quickly rolls to its critical value. This yields:
\begin{equation}
|B|=\frac{g_A^2 \xi_A + g_B^2 \xi_B}{2 \pi 
\sqrt{g_A^2 \xi_A^2 + g_B^2 \xi_B^2}} {\mathrm M}_P.
\end{equation}
We can now compare $\delta_H^2 |_A$ and $\delta_H^2 |_B$ at the end of the
first stage. It appears that for a natural choice of free
parameters ($g_A \sim g_B$, $\xi_A \sim \xi_B \sim |\xi_A - \xi_B|$),
both terms can give a dominant contribution to the global primordial
spectrum.
If $\delta_H^2 |_B$ dominates, we expect $p>1$, and a
tilted large-scale power spectrum
(due to the decrease of $|B|$, which is at the end of its 
slow-roll stage). On the other hand, if $\delta_H^2 |_B$ is
negligible, the step amplitude is directly
given by (\ref{p}), and the large
scale plateau is approximately flat, with $n_S \simeq 0.98$ 
like in single-field D-term inflation. As we said in the introduction,
this latter case is the most interesting one
in the framework of $\Lambda$CDM models.
A numerical study shows that $\delta_H^2 |_A \gg \delta_H^2 |_B$
holds in a wide region in parameter space. 
Indeed, we explored systematically the region
in which $0.1 \leq g_B/g_A \leq 10$, and $(\sqrt{\xi_A}, \sqrt{\xi_B})$ 
are in the range $ (0.1-10) \times 10^{-3} {\mathrm M}_P$. 
We find that $\delta_H^2 |_A \geq 10~\delta_H^2 |_B$ whenever:
\begin{equation} 
(g_B \geq 0.8 g_A,~\sqrt{\xi_A} \geq 1.1 \sqrt{\xi_B}) \quad 
{\rm or} \quad 
(\forall g_A, \forall g_B,~\sqrt{\xi_B} \geq 1.1 \sqrt{ \xi_A}).
\end{equation}
So, inside these two regions, the primordial
spectrum has two approximately scale-invariant plateaus $(n_S \simeq 0.98)$, 
and the step amplitude is given by (\ref{p}).
Further, a good agreement with observations requires a small inverted step,
$0.75 \leq p \leq 0.85$ \cite{LPS1}, and of course, a correct order of
magnitude for the amplitude, 
$\delta_H^2(k_1) \sim {\cal O} (10^{-10})$. These additional constraints
single out two regions in parameter space:
\begin{eqnarray}
&& \left( 2.2 g_A \leq g_B \leq 4 g_A,
~13(\sqrt{\xi_A}-4.5 \! \times \! 10^{-3}) 
\leq 10^3 \xi_B
\leq 8(\sqrt{\xi_A}-2.5 \! \times \! 10^{-3}) \right)
\nonumber \\
&{\rm or}&
\quad
\left( g_B \leq 1.5 g_A,
~150\xi_A-2.5 \! \times \! 10^{-3}
\leq \sqrt{\xi_B}
\leq 90\xi_A-4.3 \! \times \! 10^{-3} \right).
\end{eqnarray}

In this section, we only studied the primordial spectrum of
adiabatic perturbations. Indeed, it is easy to show that tensor 
contributions to large-scale CMB anisotropies are negligible in this model, 
like in usual single-field D-term inflation (a significant
tensor contribution would require $g_{A,B} \sim {\cal O} (10)$ 
or greater, while the consitency of the underlying SUSY or SUGRA theory 
requires $g_{A,B} \sim {\cal O} (10^{-1})$ or smaller \cite{HA2}). 

\subsection{Transition between slow-roll inflationary stages}

We will briefly discuss the issue of primordial spectrum calculation
for $k_1 < k < k_2$. During this stage, $A$ is still slow-rolling, but
$B$ and $B_-$ obey to the second-order differential equations in global
susy:
\begin{equation} \label{motion}
\ddot{\tilde{B}}_{(-)}+3H \dot{\tilde{B}}_{(-)}+
\frac{\partial V}{\partial \tilde{B}_{(-)}}=0.
\end{equation}
The derivatives of the potential are given by the tree-level
expression (\ref{vtree}), 
plus complicated one-loop corrections.
In supergravity, the tree-level potential will be different from 
(\ref{vtree}), even with a minimal gauge kinetic function
and K\"ahler potential. With a non-minimal K\"ahler potential, an additional 
factor will also multiply both inertial and damping terms
$\ddot{\tilde{B}}_{(-)}+3H \dot{\tilde{B}}_{(-)}$.
The evolution of background quantities has been
already studied in a similar situation in \cite{LT}.

As far as perturbations are concerned, the simplest possibility would be to
recover the generic primordial power spectrum introduced by Starobinsky 
\cite{S92}, also in the case of a transition between 
two slow-roll inflationary stages, with a jump in the potential derivative. 
This would be possible if :
(i) $H$ was approximately constant in the vicinity of the 
transition; (ii) the number of {\it e}-folds 
separating both slow-roll regimes
was much smaller than one (in other terms, the time-scale of the transition 
must be much smaller than $H^{-1}$). However, in the model under 
consideration, (i) is
not a good approximation, because during the transition there is a jump
in the potential itself. Moreover, (ii) is in contradiction with the
equations of motion (\ref{motion}) : $B$ and $B_-$ only stabilize after 
a time-scale $\delta t \sim H^{-1}$, due to
the damping factor $+3H$. This statement is very general, and holds
even in supergravity with a non-minimal K\"ahler potential. So, it seems
that only a first-order phase transition can reproduce the exact power
spectrum of \cite{S92}, 
as previously noticed by Starobinsky himself \cite{S98}.
A model of inflation with a first-order phase transition has been proposed
in \cite{BAFO}.

Further, we don't believe that the power spectrum
in our model can be approximated by the numerical solution of 
double chaotic inflation \cite{P94}. 
Indeed, in chaotic double inflation, fields masses are typically of the
same order as the Hubble parameter: so, background fields have no strong
oscillatory regime. In our case, estimating the effective mass for $B_-$
from the tree-level potential, and comparing it with 
$V$ (which gives a lower bound on $H$), 
we see that before stabilization, $B_-$
will undergo approximately $\sim {\mathrm M}_P/\sqrt{\xi_B} \sim 10^3$
oscillations. These oscillations are relevant for the primordial spectrum
calculation, because $B_-$ perturbations and metric perturbations will
strongly couple.

So, for each particular model, a 
numerical integration of the background and perturbation equations should
be performed in order to find the shape, and even the width of the
transition in the primordial power spectrum. 
The result could be either a step or a bump, and since
the fields stabilize in $\delta t \sim H^{-1}$, it is reasonable to 
think that the width of the 
feature will not be too large, $(k_2/k_1)\leq 10$. So, in any case, the 
primordial spectrum should be in good agreement with current observations,
but precise comparison with future data requires numerical work.

\section{Conclusion}

We introduced a model of double D-term inflation, with two $U(1)$ gauge
symmetries, and two associated Fayet-Iliopoulos terms. 
A phase of instability for two fields $(B, B_-)$ separates two slow-roll
inflationary stages. During this transition, $B_-$ partially cancels the
Fayet-Iliopoulos terms, causing a jump in the potential; also, since 
it is charged under both symmetries, it affects the one-loop corrections 
in such way that the potential can become less steep in the direction of one
inflaton field, $\tilde{A}$. As a result, for a wide window in the space
of free parameters  
(the Fayet-Iliopoulos terms $\xi_{A,B}$ and the gauge
coupling constants $g_{A,B}$), 
the primordial spectrum of adiabatic perturbations consists
in two approximately scale-invariant plateaus, separated by an unknown
feature, presumably of small width, and with {\it more} power on small scales.
The amplitude of the step, $p$, is given by a simple function of parameters 
(\ref{p}). In the framework of $\Lambda$CDM, spectra with
$0.75 \leq p \leq 0.85 $ are likely to
fit very well current LSS and CMB data, as argued in \cite{LPS1,LPS2}. 
However, for a detailed comparison of our model
with observations, we need the shape of the primordial power spectrum 
between the plateaus. This issue requires a numerical
integration, and the result will be model-dependent, in contrast with 
predictions for the slow-roll plateaus. 
Finally, before any precise comparison with observations, 
one should also consider
the production of local cosmic strings, which is a typical 
feature of D-term inflation \cite{J}. Indeed, CMB anisotropies
and LSS may result from both local cosmic strings and inflationary
perturbations \cite{CHM}.

\section*{Acknowledgements}
I would like to thank D.~Polarski, A.~Starobinsky and N.~Tetradis for 
illuminating discussions. G.~Dvali, E.~Gawiser and R.~Jeannerot also
provided very useful comments on this work. I am supported by the
European Community under TMR network contract No. FMRX-CT96-0090.


\begin{thebibliography}{99}
\bibitem{LL}A.~Liddle and D.~Lyth,
Phys. Rept. 231 (1993) 1.
\bibitem{KLS}L.~Kofman, A.~Linde and A.~Starobinsky,
Phys. Lett. B 157 (1985) 361; \\
L.~Kofman and A.~Linde, Nucl. Phys. B 282 (1987) 555; \\
J.~Silk and M.~Turner, Phys. Rev. D 35 (1987) 419; \\
L.~Kofman and D.~Pogosyan, Phys. Lett. B 214 (1988) 508; \\
S.~Gottl\"{o}ber, V.~M\"{u}ller and A.~Starobinsky, Phys. Rev. D 43 (1991) 
2510; \\ 
D.~Polarski and A.~Starobinsky, Nucl. Phys. B 385 (1992) 623. 
\bibitem{RS}J.~Adams, G.~Ross and S.~Sarkar, Phys. Lett. B 391 (1997) 271;
Nucl. Phys. B 503 (1997) 405.
\bibitem{ST}M.~Sakellariadou and N.~Tetradis, hep-ph/9806461.
\bibitem{P94}D.~Polarski, Phys. Rev. D 49 (1994) 6319.
\bibitem{S92}A.~Starobinsky, JETP Lett. 55 (1992) 489.
\bibitem{LP}J.~Lesgourgues and D.~Polarski, Phys. Rev. D 56 (1997) 6425.
\bibitem{LPS1}J.~Lesgourgues, D.~Polarski and A.~Starobinsky, 
MNRAS 297 (1998) 769.
\bibitem{LPS2}J.~Lesgourgues, D.~Polarski and A.~Starobinsky,
astro-ph/9807019.
\bibitem{PGRW}S.~Perlmutter, G.~Aldering, N.~Della Valle, S.~Deutsua et al.,
Nature 391 (1998) 51; \\
P.~Garnavitch, R.~Kirshner, P.~Challis, J.~Tonry et al., ApJ Lett.
(1998) 493, L53; \\
A.~Riess, A.~Filippenko, P.~Challis, A.~Clocchiatta et al., 
astro-ph/9805201; \\
M.~White, ApJ 56 (1998) 495.
\bibitem{GB} E.~Gazta\~naga and C.~Baugh, MNRAS 294 (1998) 229.
\bibitem{E}J.~Einasto, M.~Einasto, S.~Gottl\"{o}ber, V.~M\"{u}ller 
et al., Nature 385 (1997) 139; \\
J.~Einasto, M.~Einasto, P.~Frisch, S.~Gottl\"{o}ber 
et al., MNRAS 289 (1997) 801; \\
J.~Einasto, M.~Einasto, P.~Frisch, S.~Gottl\"{o}ber 
et al., MNRAS 289 (1997) 813; \\
J.~Retzlaff, S.~Borgani, S.~Gottl\"{o}ber, A.~Klypin and V.~M\"{u}ller~V., 
New A (1998) in press, astro-ph/9709044.
\bibitem{GAS}E.~Gawiser, L.~Amendola and J.~Silk, in preparation.
\bibitem{BAFO}C.~Baccigalupi, L.~Amendola, P.~Fortini and F.~Occhionero,
Phys.Rev. D 56 (1997) 4610.
\bibitem{LLL}A.~Linde, Phys. Lett. B 259 (1991) 38; \\
E.~Copeland, A.~Liddle, D.~Lyth, E.~Stewart and D.~Wands,
Phys. Rev. D 49 (1994) 6410.
\bibitem{LR}D.~Lyth and A.~Riotto, 
Phys. Rep. (1998) in press, hep-ph/9807278.
\bibitem{BD}P.~Bin\'etruy and G.~Dvali, Phys. Lett. B 388 (1996) 241,
hep-ph/9606342.
\bibitem{HA}E.~Haylo, Phys. Lett. B 387 (1996) 43, hep-ph/9606423.
\bibitem{DR}G.~Dvali and A.~Riotto, Phys. Lett. B 417 (1998) 20.
\bibitem{ERR}J.~Espinosa, A.~Riotto and G.~Ross, Nucl. Phys. B 531 (1998)
461.
\bibitem{HA2}E.~Haylo, hep-ph/9901302.
\bibitem{LT}G.~Lazarides and N.~Tetradis, Phys. Rev. D 58 (1998) 123502.
\bibitem{BCT}Z.~Berezhiani, D.~Comelli and N.~Tetradis, 
Phys.Lett. B 431 (1998) 286.
\bibitem{DSS}G.~Dvali, Q.~Shafi and R.~Schaefer, 
Phys. Rev. Lett. 73 (1994) 1886.
\bibitem{J}R.~Jeannerot, Phys. Rev. D 53 (1996) 5426; Phys. Rev. D 56 (1997)
6205. 
\bibitem{PS}D.~Polarski and A.~Starobinsky, Nucl. Phys. B 385 (1992) 623; 
Phys. Rev. D 50 (1994) 6123.
\bibitem{PARA}B.~A.~Bassett, D.~I.~Kaiser and R.~Marteens, hep-ph/9808404; \\
A.~Taruya, hep-ph/9812342; \\
B.~A.~Bassett, F.~Tamburini, D.~I.~Kaiser and R.~Marteens, hep-ph/9901319.
\bibitem{S98}A.~Starobinsky, in : Large Scale Structure: 
Tracks and Traces, Proc. of the 12th Potsdam Cosmology Workshop 
(eds. V. M\"uller et al., Singapore, World Scientific, 1998).
\bibitem{CHM}C.~Contaldi, M.~Hindmarsh and J.~Magueijo, astro-ph/9809053.

\end{thebibliography}
\end{document}